\documentclass[aps,prl,twocolumn,superscriptaddress,showpacs]{revtex4}
\usepackage{graphicx}
\usepackage{xcolor}
\usepackage{amsmath,amssymb,latexsym}
\usepackage{hyperref}
\usepackage{ulem}

\begin{document}

\title{Constraints on cosmic curvature with lensing time delays and gravitational waves}
\author{Kai Liao}
\affiliation{School of Science, Wuhan University of Technology, Wuhan 430070, China}

\begin{abstract}
 Assuming the $\Lambda$CDM model, the CMB and BAO observations indicate a very flat Universe.
  Model-independent measurements are therefore worth studying.
  Time delays measured in lensed quasars provide the time delay distances. When compared with the luminosity
  distances from Supernova Ia observation, the measurements can provide the curvature information under the Distance Sum Rule of FLRW metric.
  This method is limited by the low redshifts of SNe Ia. In this work, we propose gravitational waves from the Einstein Telescope as standard
  sirens which reach higher redshifts covering
  the redshift range of lensed quasars from Large Synoptic Survey Telescope, could provide much more stringent constraints on the curvature.
  We first consider a conservative case where only 100 gravitational waves with electromagnetic counterparts are available,
  the $1\sigma$ uncertainty for the curvature parameter $\Omega_k$ is 0.057. In an optimistic case with 1000 signals available, then $\Omega_k$ uncertainty is 0.027.
  Combining with SNe Ia from Dark Energy Survey, $\Omega_k$ can be further constrained to 0.027 and 0.018, respectively.

\end{abstract}

\maketitle

\section{Introduction}
$\Lambda$CDM is generally considered as the best description of the Universe, where $\sim 70\%$
of the energy budgets come from the dark energy mimicking a cosmological constant and accelerating the Universe,
while most of the rest are from cold dark matter. Recently, this concordance scenario has been challenged.
The local supernova Ia (SN Ia) observation based on distance ladder method claims an
obviously larger Hubble constant ($H_0$) than the cosmic microwave background measurements based on $\Lambda$CDM model, i.e.,
$H_0$ has been unprecedentedly constrained, however, in different directions~\cite{Freedman2017}.
This ``$H_0$ tension problem" could arise from either systematics or a violation of $\Lambda$CDM. Development of independent probes, for example,
the strong lensing time delays~\cite{H0LiCOW,Birrer2018}, may help us understand this discrepancy or reveal new physics.

The cosmic curvature is another crucial cosmological parameter which affects the evolution of our Universe and the dark energy properties.
Any deviation from a flat Universe would bring big problems in inflation theory and fundamental physics.
The CMB plus BAO measurements have shown the Universe is quite flat~\cite{Eisenstein2005,Tegmark2006,Planck2016} with
the latest constraint $\Omega_k=0.001\pm0.002$~\cite{Planck2018}. However, like $H_0$, the results are based on
$\Lambda$CDM. Therefore, model-independent measurements of the curvature can also test $\Lambda$CDM
and are worth developing. Similarly, any curvature tension with CMB might be another evidence of $\Lambda$CDM violation.

Moreover, the curvature is related to testing the FLRW metric, a more fundamental cosmological assumption~\cite{Clarkson2008}.
By comparing observational determinations of the expansion rates and cosmological distances, the curvature evolving with redshift
was model-independently reconstructed to test the FLRW metric~\cite{Shafieloo2010,Sapone2014,Cai2016}, though the constraints were weak due to the requirement
of constructing the derivative of noisy distance measure data in these methods. To avoid this, the comoving distances were reconstructed by
Hubble parameter data and used to confront with luminosity distances~\cite{Li2016} and angular diameter distances~\cite{Yu2016} which encode the curvature.
However, the reconstructed comoving distances are correlated by Gaussian Process and may underestimate the uncertainties~\cite{Liao2015a}.

Strong gravitational lensing by galaxies~\cite{Treu2010} has become a useful tool in studying astrophysics and cosmology.
There are two popular approaches to measuring the distances. Firstly, one could assume a simple lens model,
for example, the Singular Isothermal Sphere (SIS) or its extensions, and apply to a set of lenses. With the measurements of central velocity dispersions,
the Einstein radii, the two diameter distance ratios $D_{ls}/D_s$ can be inferred. Under the Distance Sum Rule (DSR), the curvature and
be constrained by comparing with SNe Ia~\cite{Rasanen2015}. However, the uncertainty was quite large $-1.22<\Omega_k<0.63$ within
$2\sigma$ uncertainty. Besides, the systematics would be large if one takes the universal simple lens model~\cite{Xia2017,Li2018,Jiang2007}.
To get a robust constraint, the time delay method were proposed where the lens modelling processes are for individual lenses~\cite{Liao2017,Denissenya2018}.
The observation of host arcs, velocity dispersion, and AGN images with time delay measurements from their light curves can provide precise and accurate
information on the time delay distance $D_lD_s/D_{ls}$. However, the redshifts of SNe Ia from Dark Energy Survey (DES) are low $<1.3$ while
the lens sources from Large Synoptic Survey Telescope
(LSST) can reach $z\sim5$~\cite{Liao2017}. We can therefore only utilize a small fraction of the lensing data. Besides, supernovae only give the relative
distances with unknown intrinsic brightness. The direct luminosity distances with high redshifts
can significantly improve the constraint precision.

On the other hand, gravitational waves (GWs) predicted by General Relativity are transverse waves of spatial strain, travelling at speed of light.
Several signals have recently been recorded by Advanced LIGO and VIRGO detectors~\cite{GW150914,GW151226,GW170104} including a binary neutron star system with electromagnetic
counterparts~\cite{GW170817}. As standard sirens, the waveforms of chirping signals from binary star systems provide the direct luminosity distance information~\cite{Schutz1986,SS}.
If the redshifts of GW sources can be measured in other astrophysical approaches, for example, from their electromagnetic (EM) counterparts,
we can get the distance-redshift relation applied in cosmology. The
short gamma ray burst (SGRB) is one of the most promising EM counterparts, once it is confirmed, the redshift can be measured from its host galaxy or afterglow.
The next generation detectors like the Einstein Telescope (ET) will broaden the accessible volume of the Universe by three orders of magnitude promising tens to hundreds of thousands of detections per year. The detection of binary systems can reach $z\sim5$ with signal-noise-ratio
$SNR>8$~\cite{Cai2016}. Therefore, we propose with such deep observation of GWs,
the whole lensing data from LSST can be used to infer the curvature resulting in more precise constraints.

The paper is structured as follows. The lensed quasar observation from LSST and the GW observation from ET are introduced
in Section 2 and Section 3, respectively. The methodology and results are presented in
Section 4. We give summaries and discussions in Section 5. The fiducial model used in the simulation is the flat $\Lambda$CDM with $\Omega_M=0.3, H_0=70km/s/Mpc$.

\section{Lensing observation in LSST era}
\begin{figure}
  \includegraphics[width=8cm,angle=0]{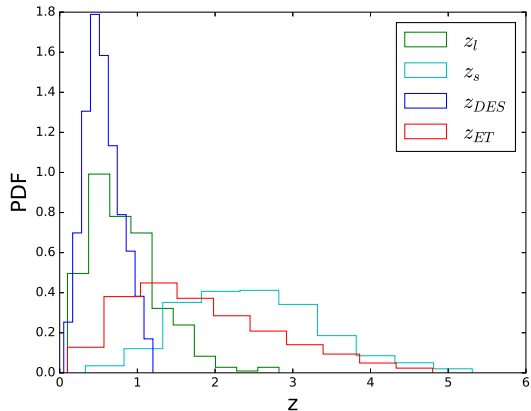}
  \caption{The redshift distributions for lensing observation including the source and the lens from LSST, the SNe Ia from DES, and the GWs from ET.
  }\label{z}
\end{figure}

\begin{figure}
  \includegraphics[width=8cm,angle=0]{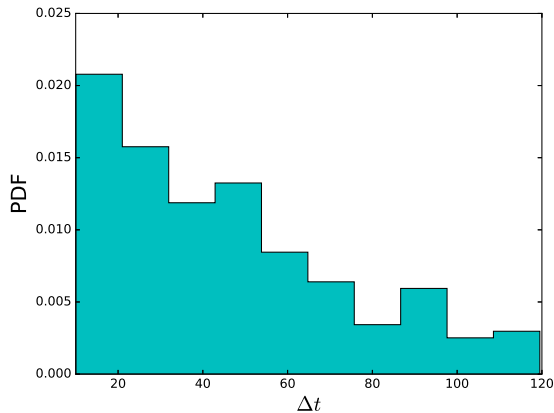}
  \caption{Distribution of time delay between AGN images for LSST lensing system. The range is chosen between 10-120 days to adjust to the LSST observation strategy.
  }\label{dt}
\end{figure}

\begin{table*}
\center
 \begin{tabular}{lccc}
  \hline\hline
  $\sigma_{\Delta t}^{LC}$& $\sigma_t^{ML}$ & $\sigma_{\Delta\phi}^{LM}/\Delta\phi$ & $\sigma^{LOS}_{\Delta\phi}/\Delta\phi$\\
  \hline
   1 day & 1 day & $3\%$ & $2.5\%$ \\
  \hline\hline
 \end{tabular}
 \caption{The lensing-related uncertainties adopted in this work.
}\label{lensinig uncertainties}
\end{table*}
The upcoming LSST will monitor nearly half of the sky for 10 years by repeatedly scanning the field.
It is supposed to find and monitor $\sim10^3$ lensed quasars~\cite{OM10}. With high-quality and long-time light curves,
the time delay between AGN images can be measured precisely and accurately. To assess the current algorithms
of time delay extraction, a time delay challenge (TDC) has been conducted~\cite{Liao2015b}. While a group generated thousands
of realistic light curve pairs, the community was invited to extract time delays using their algorithms blindly.
The results showed there will be 400 well-measured systems with average time delay precision $\sim3\%$ and accuracy $<1\%$.
The redshift and time delay distributions are plotted in Fig.\ref{z} and Fig.\ref{dt}.
The lessons in TDC showed the absolute uncertainty is approximately a constant mainly determined by the cadenced sampling.
Different from the previous work which took a constant precision for each system~\cite{Liao2017},
we take 1 day as the absolute time delay uncertainty from light curves for each system, such that the average precision is $3\%$.

Moreover, Tie $\&$ Kochanek (2018) recently proposed that the microlensing by stars in the lens galaxy could change the
actual time delay on the light-crossing time scale of the emission region $\sim days$, due to
the finite AGN accretion disc and the differential magnification of the coherent temperature fluctuations.
The microlensing time delay is an absolute error. However, a good AGN model has not been set up and some parameters of the AGN accretion may not
be observed, for example, the inclination and the accretion size. Therefore, this systematics is worth further studying by the community.
To incorporate this effect, we consider extra 1 day as the systematics from microlensing induced time delay. This value is chosen due to
the characteristic variation of the microlensing time delay map~\cite{Tie2018,Bonvin2018,Chen2018}, though current studies have not found a nonzero one~\cite{Bonvin2018,Chen2018}.
A detailed study of the impact of choosing different values (0.3-3 days) on cosmology is presented in~\cite{Li2018}.

The measured time delay is related with cosmological distances and the lens through~\cite{Treu2010,Treu2016}:
\begin{equation}
\Delta t=\frac{(1+z_l)D_{\Delta t}}{c}\Delta\phi,\label{relation}
\end{equation}
where $c$ is light speed, $D_{\Delta t}=D_lD_s/D_{ls}$ is called the ``time delay distance" which is the combination of three angular
diameter distances, subscripts ``l" and ``s" stand for the lens and the source, respectively.
$\Delta\phi$ is the Fermat potential between two images determined by the lensing potential.

On the other hand, the state-of-art lensing program H0LiCOW~\cite{H0LiCOW} gives an percent level precision on
lens modelling, specifically, $\Delta\phi$~\cite{Bonvin2017}. With the observation of host arcs by deep imaging of HST, the central velocity dispersion from spectroscopy,
and the image positions, the Fermat potential difference can be inferred under the Bayesian framework. The systematics can
be well-controlled by ``blind analysis" which is an ongoing work~\cite{Ding2018}. We take $3\%$ as the relative uncertainty from lens modelling for each system.
Besides, the light-of-sight density fluctuation could also change
the lensing potential and inversely perturb the time delay. one can measure its impact by spectroscopic/photometric observations of local galaxy groups and LOS structures in combination with ray-tracing through N-body simulations, or even realistic simulations of lens fields~\cite{Collett2013,Greene2013,McCully2014,Treu2016}.
We take $2.5\%$ relative uncertainty as in HE0435-1223~\cite{Rusu2017} in this work. The lensing uncertainties are summarized in Tab.\ref{lensinig uncertainties}.

\begin{figure}
  \includegraphics[width=8cm,angle=0]{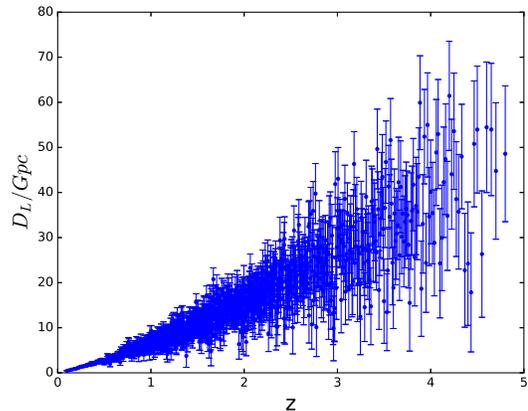}
  \caption{The simulated 1000 luminosity distances from GW standard siren observation by ET. The noise realization has been given.
  }\label{sim}
\end{figure}

\section{Standard sirens from ET}
While the SNe Ia as the standard candles measure the relative distances based on the distance ladders,
the chirping GW signals from inspiralling and merging compact binary stars are self-calibrating which directly give
the absolute luminosity distances, known as standard sirens~\cite{Schutz1986,SS}.
GW observation should also be unaffected by potential cosmic opacity like SNe Ia~\cite{Liao2015a}.
For cosmological applications, the important thing is to get the redshift information, since GW itself has the redshift-chirp mass degeneracy problem. This
can be done by different approaches, for example, using galaxy catalog as priors~\cite{Schutz1986,Fan2014},
by comparing measured (redshifted) mass distribution of NSs with a universal rest-frame NS mass distribution~\cite{Taylor2012}, or through
tidal deformation of neutron stars knowing the equation-of-state~\cite{Messenger2012,Messenger2014}. Among these, the simplest and most robust way is to find the EM
counterparts. The binary merger of neutron stars (NS-NS) or neutron star-black hole (NS-BH) is generally considered as the progenitor
of SGRBs, besides, kilonovae or even fast radio bursts can also be the EM counterparts~\cite{counterparts}. The GW170817 has
observed the EM emissions covering all frequencies~\cite{GW170817}. We can therefore measure redshifts from either the EM counterparts themselves or their the host galaxies.

While current LIGO and VIRGO detectors mainly aim at detecting binary neutron star signals nearby $z_{max}<0.1$,
the next-generation detectors like ET will reach much deeper Universe and bring us $\sim10^5$ NS-NS and NS-BH systems.
For SGRBs, they are likely to be strongly beamed, which allow inferring the inclinations of the binaries breaking
the distance-inclination degeneracy. However, we can only observe the SGRBs from the nearly face-on systems, the probability is $\sim 10^{-3}$~\cite{Cai2016}.
Therefore, we can assume to see $\sim10^2$ signals with accurate redshift measurements.

Following the simulation process by Cai $\&$ Yang (2016), we consider ET will register 100 or 1000 GWs with
redshift measurements. The NS and BH mass distributions are chosen uniformly in [1,2]$M_{\odot}$
and [3,10]$M_{\odot}$, respectively. The BH distribution follows~\cite{Fryer2011}. The ratio between BHNS and BNS events
is 0.03. The redshift distribution follows~\cite{Cai2016,Zhao2011}:
\begin{equation}
P(z)\propto \frac{4\pi \chi^2(z)R(z)}{H(z)(1+z)},
\end{equation}
where $\chi(z)$ is the comoving distance and
\begin{equation}
R(z)=\begin{cases}
1+2z, & z\leq 1 \\
\frac{3}{4}(5-z), & 1<z<5 \\
0, & z\geq 5.
\end{cases}
\label{equa:rz}
\end{equation}

For a nearly face-on case, the instrumental uncertainty is given by~\cite{Cai2016}:
\begin{align}
\sigma_{D_L}^{\rm inst}\simeq \sqrt{\left\langle\frac{\partial \mathcal H}{\partial D_L},\frac{\partial \mathcal H}{\partial d_L}\right\rangle^{-1}},
\end{align}
where the angle bracket is the inner product.
$\mathcal H \propto D_L^{-1}$ is the Fourier transform of the waveform, then $\sigma_{D_L}^{\rm inst}\simeq D_L/\rho$, where $\rho$ is the combined SNR,
determined by the square root of the inner product of $\mathcal H$. $\rho>8$ is the minimum requirement for detecting a GW signal.
Considering the inclination $\iota$, the maximal effect of the inclination on the SNR is a factor of 2 (between $\iota =0^{\circ}$ and $\iota = 90^{\circ}$)~\cite{Cai2016}.
Therefore, the luminosity distance instrumental uncertainty is set to be
\begin{align}
\sigma_{D_L}^{\rm inst}\simeq \frac{2D_L}{\rho}.
\label{sigmainst}
\end{align}
Besides, weak leasing effects by large-scale structure can bias the results especially at high redshifts. Following~\cite{Zhao2011},
we take the $\sigma_{D_L}^{wl}/D_L=0.05z$.
The total uncertainty on the luminosity distance is given by:
\begin{equation}
\sigma_{D_L}=\sqrt{(\sigma_{D_L}^{\rm inst})^2+(\sigma_{D_L}^{wl})^2}.
\end{equation}
We refer to Cai $\&$ Yang (2016) for more simulation details.
Given specific noise realization, the simulated luminosity distances from standard sirens is presented in Fig.\ref{sim}.

\section{Methodology and results}

\begin{table*}
\center
 \begin{tabular}{lcccc}
  \hline\hline
  & 100 GWs& 1000 GWs & 100 GWs+SNe Ia & 1000 GWs+SNe Ia\\
  \hline
  $\sigma_{\Omega_k}$ & 0.057 & 0.027 & 0.027 & 0.018 \\
  \hline\hline
 \end{tabular}
 \caption{$1\sigma$ uncertainties of the curvature parameter $\Omega_k$ for 100 and 1000 GWs with and without SNe Ia, respectively.
}\label{resultdata}
\end{table*}

One of the basic assumptions in cosmology is the Universe is homogeneous and isotropic
at large scales, described by FLRW metric£º
\begin{equation}
d s^2 = - c^2d t^2 + \frac{a(t)^2}{1 - K r^2} d r^2 + a(t)^2 r^2 d\Omega^2,
\end{equation}
where the constant $K$ determines the spatial curvature, the Universe is close ($K>0$), open ($K<0$) or flat ($K=0$).
The dimensionless distance between redshifts $z_l$ and $z_s$ is related with
the comoving distance $\chi=\int_{z_l}^{z_s}\frac{1}{H(z)}dz=\int_{t_s(z_s)}^{t_l(z_l)}\frac{H_0}{a(t)}dt$ by
\begin{equation}
d(z_l, z_s)=\frac{1}{\sqrt{|\Omega_k|}}\begin{cases}  \sinh\left( \sqrt{\Omega_k} \chi \right)&\Omega_k>0\\ \chi &\Omega_k=0\\ \sin\left( \sqrt{-\Omega_k} \chi \right)&\Omega_k<0,
\end{cases}
\end{equation}
where $\Omega_k=-K/H_0^2$ is the curvature parameter. The luminosity distance $D_L=c(1+z)d/H_0$ and angular diameter distance $D_A=(1+z)^2D_L$.

Denoting $d(z)\equiv d(0,z),d_{ls}\equiv d(z_l,z_s),d_l\equiv d(z_l),d_s\equiv d(z_s)$, the DSR gives:
\begin{equation}
d_{ls} = \epsilon_1 d_s \sqrt{1 + \Omega_k d_l^2} - \epsilon_2 d_l \sqrt{1 + \Omega_k d_s^2},\label{sumrule}
\end{equation}
where $\epsilon_i=\pm1$. For a one-to-one correspondence between t and z with $d\prime(z)>0$, the $\epsilon_i=1$.
Following~\cite{Liao2017}, we rewrite DSR such that the time delay distance and luminosity distances are encoded:
\begin{equation}
\frac{d_{ls}}{d_ld_s}=T(z_l)-T(z_s),\label{TDSR}
\end{equation}
where
\begin{equation}
T(z)=\sqrt{1/d(z)^2+\Omega_k}.\label{T}
\end{equation}
The left item can be got from time delay distance measurement by lensing observation while $d(z)$ in the right items is determined by luminosity distances
from GW or SNe Ia. The $H_0$ will be marginalized in our analysis.
By comparing the two observations, we can constrain the cosmic curvature parameter $\Omega_k$ in the DSR.
In principle, one needs to use two GWs or SNe Ia that have the same redshifts of the lens and source in the lensed quasar system.
However, there are always differences between the lensing redshifts and the nearest GWs or SNe Ia. One can assume the redshift difference can be
ignored if it is small enough or smooth the evolution of luminosity distances in a cosmological-model-independent way, making use of all discrete data, for example, using a polynomial~\cite{Rasanen2015} or Gaussian process~\cite{Shafieloo2012}.
In this work we parameterize the dimensionless distance as a fourth-order polynomial $d(z)=z+a_1z^2+a_2z^3+a_3z^4$ as in~\cite{Rasanen2015,Liao2017}. Increasing the order matters only when the precision
becomes higher. We simultaneously fit $d(z)$ by luminosity distance data and find the best-fit of $\Omega_k$ in the DSR. The statistical quantity is written as
\begin{equation}
\begin{split}
\chi^2=\sum_{i=1}^{n_L}\frac{(D_{\Delta t,i}-[c/H_0/(1+z_{l,i})]/[T(z_{l,i})-T(z_{s,i})])^2}{\sigma^2_{D_{\Delta t,i}}}+\\
\sum_{j=1}^{n_{GW}+n_{SN}}\frac{[(c(1+z_j)/H_0)d(z_j)-D_L^{obs}(z_j)]^2}{\sigma^2_{D_L(z_j)}}.
\end{split}\label{chi2}
\end{equation}
In the analysis, we simply assume $D_{\Delta t}$ as the observation following Gaussian distribution. However,
note that we get the uncertainty of $D_{\Delta t}$ through error propagation from the uncertainties in Table.\ref{lensinig uncertainties} based on Eq.\ref{relation}, it is
usually not a Gaussian distribution. In realistic cases, one should get the uncertainty of $D_{\Delta t}$ in the Bayesian framework and take the direct observations as Gaussian distributions (see the H0LiCOW program~\cite{Bonvin2017}).
In this work, firstly the results of H0LiCOW based on a full Bayesian framework indeed give
approximate Gaussian-like posterior distributions of the time delay distances~\cite{Bonvin2017}.
Secondly, we do not aim at giving an accurate constraint on curvature from realistic
data, but give an estimate of the precision level for the future observation. Therefore, while we can keep the analysis process simple and clear for the readers,
this assumption would not affect our main conclusion. More detailed discussions can be found in~\cite{Liao2019}.

\begin{figure*}
 \centering
  \includegraphics[width=8cm,angle=0]{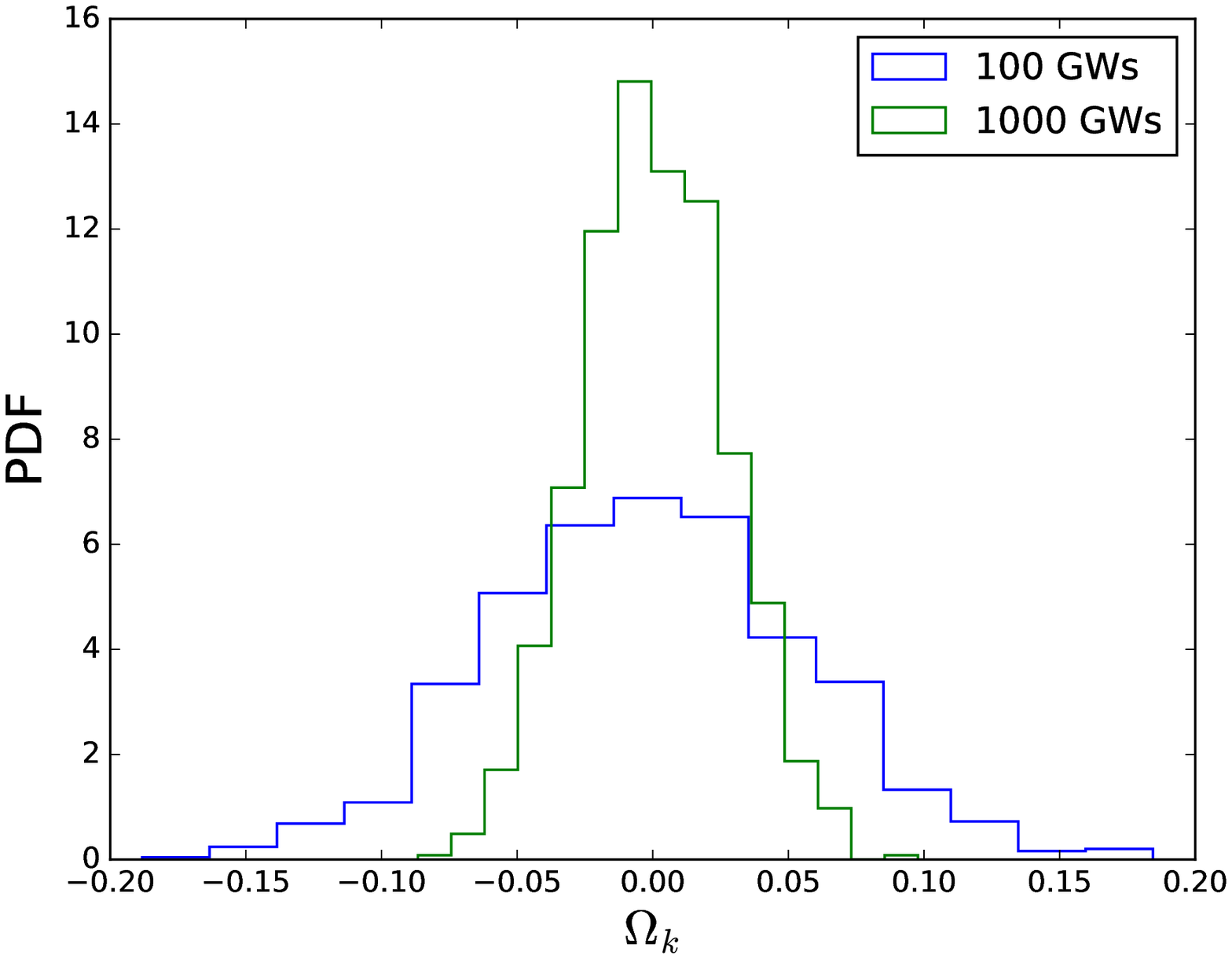}
  \includegraphics[width=8cm,angle=0]{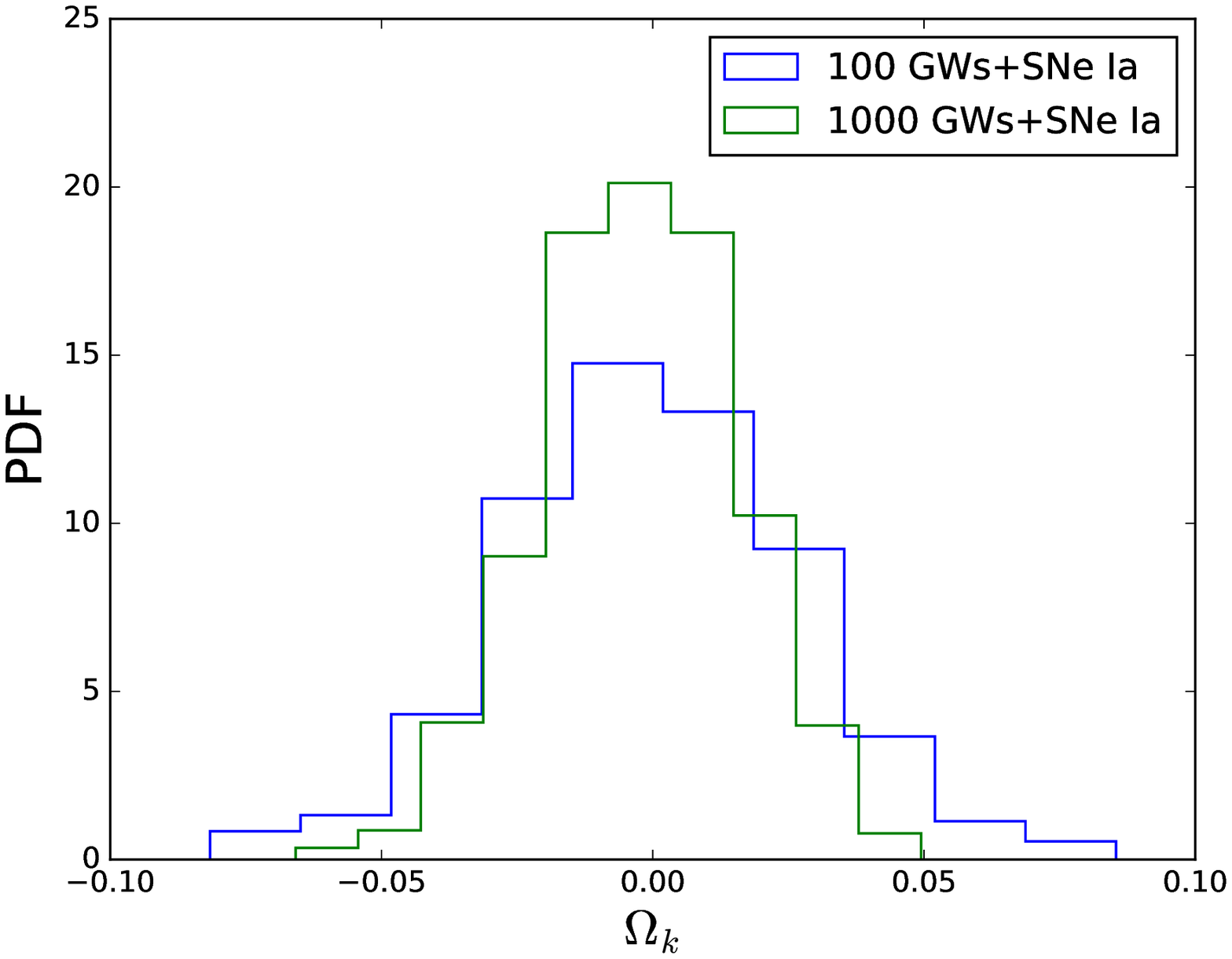}
  \caption{1-D marginalized PDFs of the curvature parameter $\Omega_k$. Left: 100 and 1000 GWs
  by ET with redshift measurements. Right: results for combining with SNa Ia observation by DES.
  }\label{result}
\end{figure*}

As a method and prediction work, we can not apply a direct Bayesian analysis based on Eq.\ref{chi2} from observed data.
To make a prediction of constraints on $\Omega_k$, we use minimization statistics~\cite{Liao2019}.
We first generate 3000 sets of simulated data under different noise realizations, then
for each dataset, we do minimizations to find the best-fit values of all free parameters including the coefficients of the polynomial, $H_0$ and $\Omega_k$.
Lastly, we study all the best-fit values
of $\Omega_k$ and plot the PDFs in Fig.\ref{result} as the marginalized distributions. We take the PDFs approximately as Gaussian distributions (see Fig.\ref{result})
and adopt the standard deviations as the $1\sigma$ uncertainties, the statistic results are summarized in Tab.\ref{resultdata}.
As the results show, for only 100 GWs, $\Omega_k$ can be constrained to 0.057, comparable with previous SNe Ia from DES.
For 1000 GWs, the constraint would be 0.027, much stringent then SNe Ia. We also combine 3443 SNe Ia from DES to improve the constraint for
$d(z)$, then results are 0.027 and 0.018, for 100 and 1000 GWs, respectively.

\section{Summaries and Discussions}
Inspired by the previous work where high-redshift and direct luminosity distance observation was found to be crucial for constraining the cosmic curvature,
we propose the GW observation by the third-generation detectors like ET would significantly improve the result.
By simulating more realistic time delay measurements in the lensing data and using simulated luminosity distances from GWs plus SNe Ia,
the constraint power would be at least doubly strengthened.

Our method is cosmological-model-independent based on the Distance Sum Rule in FLRW metric. This pure geometric test would not only give
the curvature, but also shed on light the validation of $\Lambda$CDM model where the Universe is quite flat. Moreover, the test is also related
with the test of FLRW metric since any large $\chi^2/d.o.f$ would be a sign of the deviation of DSR. Another way to test FLRW is
to introduce an evolution of curvature $\Omega_k(z)$ to see whether it keeps a constant~\cite{Liao2017}.

Furthermore, to make the results more robust, the systematics in the observations are worth further studying. For lensing data, the microlensing induced
time delays depend on the assumptions in the AGN model, the size and the inclination of the disc, and
the local properties of the lens at each image. The lens modelling may also suffer from the systematics~\cite{Schneider2013,Birrer2016}
which may dominate over the statistical uncertainties. One may doubt if we have already hit the systematic floor currently.
There are known cases where the large-scale substructure is in the form of discs in the lens galaxies.
The dark matter subhalos may also play an important role as satellites. We anticipate the ongoing Time Delay Lens Modelling program~\cite{Ding2018} can give an
estimate of the systematics. The 400 lensing systems we assumed may be idealized, in reality, the host galaxies might be too faint to be observed.
The results would be scaled by $\sqrt{N}$.
For GW data, using multi-messengers to deal with the degeneracy of luminosity distance with other parameters is important. Besides,
systematic errors may come from detector calibration, weak lensing, GW model uncertainties such as in the waveform-modelling and the templates.
With more GW observations in current and next generation detectors and the improvement of algorithms, systematics would be better studied and controlled.

At last, since in this work we only consider the case where the EM counterpart SGRB can be observed, for future studies, one may get the redshift information by means
of statistical methods or directly from the waveform considering the tidal deformation.
In addition, for high redshift luminosity distance measurements, quasar itself was recently proposed to be standard candle~\cite{Risaliti2018}, the distances are
estimated from their X-ray and ultraviolet emission, the redshift can be up to 5.
We anticipate this technique would be mature to get precise distance measurements.

\section*{Acknowledgments}
The author thanks Tao Yang, Xuheng Ding, Zhengxiang Li and Xi-Long Fan for helpful discussions.
This work was supported by the National Natural Science Foundation of China (NSFC) No. 11603015
and the Fundamental Research Funds for the Central Universities (WUT:2018IB012).

%\bibliography{prl_cosmology.bib}

\begin{thebibliography}{}
\bibitem{Freedman2017} Freedman, W. L. 2017, Nature Astronomy, 1, 0169
\bibitem{Birrer2018} Birrer S., Treu T., Rusu C. E., et al. 2018, arXiv:1809.01274
\bibitem{H0LiCOW} Suyu, S. H., Bonvin, V., Courbin, F., et al. 2017, MNRAS, 468, 2590
\bibitem{Eisenstein2005} Eisenstein, D. J., Zehavi, I., Hogg, D.W., et al. 2005, ApJ, 633, 560
\bibitem{Planck2016} Planck Collaboration, Ade, P. A. R., Aghanim, N., et al. 2016, A\&A, 594, A13
\bibitem{Tegmark2006} Tegmark, M., Eisenstein, D., Strauss, M., et al. 2006, PhRvD, 74, 123507
\bibitem{Planck2018} Planck Collaboration, Aghanim, N., Akrami, Y., Ashdown, M., et al. 2018, arXiv:1807.06209
\bibitem{Clarkson2008} Clarkson, C., Bassett, B. A., Lu, T. C. 2008, PhRvL, 101, 011301
\bibitem{Cai2016} Cai, R.-G., Guo, Z.-K., Yang, T. 2016, PhRvD, 93, 043517
\bibitem{Shafieloo2010} Shafieloo, A., Clarkson, C. 2010, PhRvD, 81, 083537
\bibitem{Sapone2014} Sapone, D., Majerotto, E., Nesseris, S. 2014, PhRvD, 90, 023012
\bibitem{Li2016} Li, Z., Wang, G.-J., Liao, K., Zhu, Z.-H. 2016, ApJ, 833, 240
\bibitem{Yu2016} Yu, H., Wang, F. Y. 2016, ApJ, 828, 85
\bibitem{Liao2015a} Liao, K., Avgoustidis, A., Li, Z. 2015b, PhRvD, 92, 123539
\bibitem{Treu2010} Treu T. 2010, Annu. Rev. Astron. Astrophys., 48, 87
\bibitem{Rasanen2015} R\"{a}s\"{a}nen, S., Bolejko, K., Finoguenov, A. 2015, PhRvL, 115, 101301
\bibitem{Shafieloo2012} Shafieloo, A., Kim, A. G., Linder, E. V., 2012, PhRvD, 85, 123530
\bibitem{Li2018} Li, Z., Ding, X., Wang, G.-J., Liao, K., Zhu, Z.-H. 2018, ApJ, 854, 146
\bibitem{Xia2017} Xia, J.-Q., Yu, H., Wang, G.-J., et al. 2017, ApJ, 834, 75
\bibitem{Jiang2007} Jiang G., Kochanek C. S., 2007, ApJ, 671, 1568
\bibitem{Liao2017} Liao, K., Li, Z., Wang, G.-J., Fan, X.-L. 2017, ApJ, 839, 70
\bibitem{Denissenya2018} Denissenya M., Linder, E. V., Shafieloo A., 2018, JCAP, 1803, 041
\bibitem{GW150914} Abbott, B. P., Abbott, R., Abbott, T. D., et al. 2016a, PhRvL, 116, 061102
\bibitem{GW151226} Abbott, B. P., Abbott, R., Abbott, T. D., et al. 2016b, PhRvL, 116, 241103
\bibitem{GW170104} Abbott, B. P., Abbott, R., Abbott, T. D., et al. 2017a, PhRvL, 118, 221101
\bibitem{GW170817} Abbott, B. P., Abbott, R., Abbott, T. D., et al. 2017b, PhRvL, 119, 161101
\bibitem{SS} Abbott, B. P., Abbott, R., Adhikari, R. X., et al. 2017c, Nature, 551, 85
\bibitem{Schutz1986} Schutz, B. F. 1986, Nature, 323, 310
\bibitem{OM10} Oguri M., Marshall P. J. 2010, MNRAS, 405, 2579
\bibitem{Liao2015b} Liao, K., Treu, T., Marshall, P., et. al. 2015, ApJ, 800, 11
\bibitem{Bonvin2018} Bonvin V., et al., 2018, A\&A, 616, A183
\bibitem{Chen2018} Chen G. C.-F., et al., 2018, arXiv:1804.09390
\bibitem{Liao2019} Liao K., 2019, ApJ, 871, 113
\bibitem{Tie2018} Tie S. S., Kochanek C. S., 2018, MNRAS, 473, 80
\bibitem{Treu2016} Treu T., Marshall P. J., 2016, Astron. Astropys. Rev., 24, 11
\bibitem{Bonvin2017} Bovin V., et al., 2017, MNRAS, 465, 4914
\bibitem{Ding2018} Ding, X., Treu, T., Shajib, A. J., et. al. 2018, arXiv: 1801.01506
\bibitem{Collett2013} Collett T. E., et al., 2013, MNRAS, 432, 679
\bibitem{Greene2013} Greene Z. S., et al., 2013, ApJ, 768, 39
\bibitem{McCully2014} McCully C., Keeton C. R., Wong K. C., Zabludoff A. I., 2014, MNRAS, 443, 3631
\bibitem{Rusu2017} Rusu C. E., et al., 2017, MNRAS, 467, 4220
\bibitem{counterparts} Fern\"{a}ndez R., Metzger B. D., 2016, Annu. Rev. Nucl. Part. Sci., 66, 23
\bibitem{Fan2014} Fan X., Messenger C., Heng I. S., 2014, ApJ, 759, 43
\bibitem{Taylor2012} Taylor S. R., Gair J. R., 2012, PhRvD, 86, 023502
\bibitem{Messenger2012} Messenger C., Read J., 2012, PhRvL, 108, 091101
\bibitem{Messenger2014} Messenger C., Takami K., Gossan S., Rezzolla L., Sathyaprakash B. S., 2014, PhRvX, 4, 041004
\bibitem{Fryer2011} Fryer C. L., Kalogera V., 2001, ApJ, 554, 548
\bibitem{Zhao2011} Zhao W., Van Den Broeck C., Baskaran D., Li T. G. F., 2011, PhRvD, 83, 023005
\bibitem{Birrer2016} Birrer S., Amara A., Refregier A., 2016, JCAP, 8, 020
\bibitem{Schneider2013} Schneider P., Sluse D., 2013, AAP, 559, A37
\bibitem{Risaliti2018} Risaliti G., Lusso E., 2018, arXiv:1811.02590, to appear in Nature Astronomy.



\end{thebibliography}
\end{document}